\documentclass[11pt]{amsart}
\usepackage{geometry}                
\usepackage{graphicx}
\usepackage{amssymb}
\usepackage{epstopdf}
\title{Compressive Sampling for the Packet Loss Recovery in Audio Multimedia Streaming}
\author{Angelo Ciaramella and Giulio Giunta \\  Dept. of Science and Technology, Isola C4, Centro Direzionale, I-80143, Napoli (NA), ITALY; email: {angelo.ciaramella, giulio.giunta}@uniparthenope.it}

\begin{document}
\begin{abstract}
The aim of this paper is to introduce a new schema, based on a Compressive Sampling technique, for the recovery of lost data in multimedia streaming. The audio streaming data are encapsuled in different packets by using an interleaving technique. The Compressive Sampling technique is used to recover audio information in case of lost packets. Experimental results are presented on speech and musical audio signals to illustrate the performances and the capabilities of the proposed methodology.
\end{abstract}

\maketitle

\section{Introduction}
Streaming technologies and increased bandwidth in access networks have facilitated the transmission of multimedia
content on the Internet \cite{Pozueko}. This new service makes possible, for example, Internet TV or audio/video services on demand, which in
turn create great interest in various fields. Users are increasingly turning to this type of services and providers try to offer
better quality to meet such needs. The main limitation of this technology is the need for stable transmission conditions to guarantee a certain degree of Quality of Service (QoS). The development of IP multicast and the Internet multicast backbone (Mbone) has led to the emergence of a new class of scalable audio/video conferencing applications. Factors such as packet loss, delay and network congestion directly affect the quality of audio and video \cite{Pozueko,Perkins}. Several works examine the loss characteristics of such an IP multicast channel and how these affect audio communication, and a number of techniques for recovery from packet loss on the channel are studied and proposed \cite{Banu,Perkins,Feamster,Lu}.

\noindent From the other hand, over the last few years, an alternative sampling/sensing theory, known as ``Compressive Sampling'' or ``Compressed Sensing'' , enables the faithful recovery of signals, images, and other data from what appear to be highly sub-Nyquist-rate samples \cite{Candes}.  Most signals are sparse or compressible in the sense that they can be encoded with just a few numbers without much numerical or perceptual loss. Moreover, useful information content in compressible signals can be captured via sampling or sensing protocols that directly condense signals into a small amount of data. To recovery the signals we use an optimization approach based on a $L_1$ norm \cite{Candes,Romberg}. There are, however, other algorithmic approaches to Compressive Sampling based on greedy algorithms such as Orthogonal Matching Pursuit \cite{Mallat,Tropp}, Iterative Thresholding \cite{Fornasier}, Compressive Sampling Matching Pursuit \cite{Needell}, and many others.

\noindent In this paper we propose a new schema for data loss recovery in audio streaming. In the streaming model, the audio data are encapsuled in different packets by using an interleaving technique and information of the lost packets is recovered by using a Compressive Sampling technique.

\noindent The paper is organized as follows. In Section \ref{sec_1} some aspects of the streaming and lost packets are introduced. The Compressive Sampling methodology is presented in Section \ref{sec_2}. In Section \ref{sec_3} and Section  \ref{sec_4} we present the proposed methodology and some experimental results, respectively. Finally in Section  \ref{sec_5} some conclusions and future remarks are provided.

\section{Real Time Protocol and Loss Packets} \label{sec_1}
\begin{figure}[tp]
\centering
\includegraphics[width=4in,keepaspectratio]{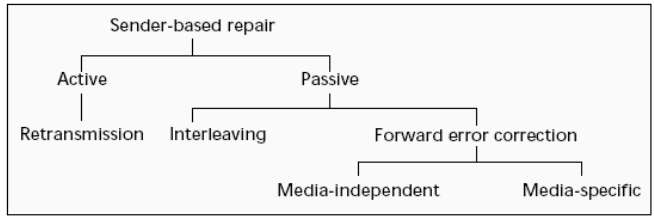}
\caption{A taxonomy of sender-based repair techniques.}
\label{fig_1}
\end{figure}

Multimedia applications require services that differ substantially from the standard ones. These applications are particularly sensitive to the end-to-end delay and they can tolerate only occasional loss of data. The concept of IP multicast to provide a scalable and efficient means by which datagrams may be distributed to a group of receivers. Internet applications, based on IP multicast, typically employ an application-level protocol to provide approximate information to the set of receivers and reception quality statistics. This protocol is the Real-time Transport Protocol (RTP) \cite{Schulzrinne}. The portion of the Internet which supports IP multicast is known as the Mbone. Multicast traffic typically
shares links with other traffic and a number of attempts have been made to characterize the loss patterns seen on the
Mbone \cite{Handley,Perkins}.  As is most clearly illustrated in \cite{Handley}, which tracks RTP reception report statistics for a large multicast session over several days, the overwhelming cause of loss is due to congestion at routers. A multicast channel will typically have relatively high latency, and the variation in end-to-end delay may be large. The delay variation is a reason for concern when developing loss-tolerant
real-time applications, since packets delayed too long will have to be discarded in order to meet the applications timing requirements, leading to the appearance of loss. This problem is more acute for interactive applications (e.g. voice over ip, conferences, wireless streaming communications).  There are a number of techniques which require the participation of the sender of an audio stream to achieve recovery from packet loss. These techniques may be split into two major classes: active retransmission and passive channel coding. It is further possible to subdivide the set of channel coding techniques, with traditional forward error correction (FEC) and interleaving-based schemes being used (see Figure \ref{fig_1} for the taxonomy). In our methodology we propose to use an interleaving-based schema.

\subsection{Interleaving}
\begin{figure}[tp]
\centering
\includegraphics[width=5in,keepaspectratio]{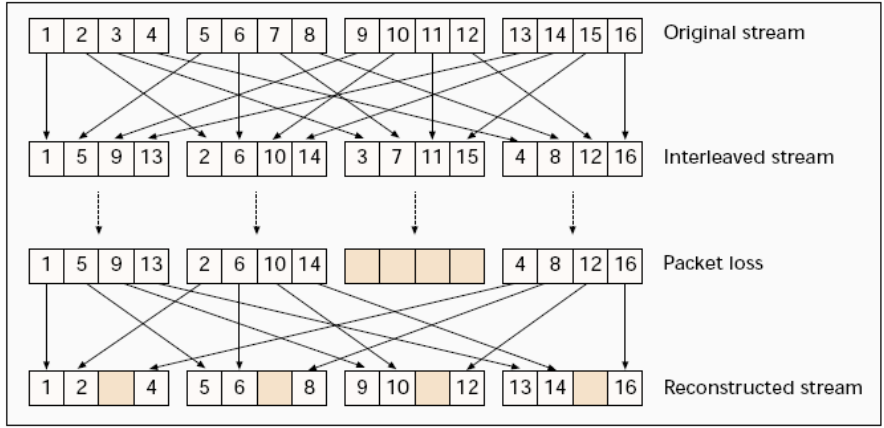}
\caption{Interleaving units across multiple packets.}
\label{fig_2}
\end{figure}

The interleaving can significantly improve the quality with which we perceive one audio stream \cite{Ramsey,Perkins}. Frames of audio signals are resequenced in packets before transmission so that originally adjacent frames are separated by a guaranteed distance in the transmitted stream and returned to their original order at the receiver. Interleaving disperses the effect of packet losses. If, for example, frames are $5$ ms in length and packets $20$ ms (i.e., $4$ frames/packet), then the first packet would contain units $1$, $5$, $9$, $13$; the second units $2$, $6$, $10$, $14$; and so on, as illustrated in Figure \ref{fig_2}. It can be seen that the loss of a single packet from an interleaved stream results in multiple small gaps in the reconstructed stream, as opposed to the single large gap which would occur in a noninterleaved stream. This spreading of the loss is important for two similar reasons: first, Mbone audio tools typically transmit packets which are similar in length to phonemes in human speech. Loss of a single packet will therefore have a large effect on the intelligibility of speech. If the loss is spread out so that small parts of several phonemes are lost, it becomes easier for listeners to mentally patch over this loss  \cite{Miller}, resulting in improved perceived quality for a given loss rate. In a somewhat similar manner, error concealment techniques perform significantly better with small gaps, since the amount of change in the signals characteristics is likely to be smaller. The majority of speech and audio coding schemes can have their output interleaved and may be modified to improve the effectiveness of interleaving. The disadvantage of interleaving is that it increases latency. The major advantage of interleaving is that it does not increase the bandwidth requirements of a stream.

\section{Compressive Sensing} \label{sec_2}
Compressive Sensing (CS) or Compressed Sensing theory asserts that one can recover certain signals from far fewer samples or measurements than traditional methods use \cite{Candes,Donoho}. To make this possible, CS relies on two principles: {\it sparsity}, which pertains to the signals of interest, and {\it incoherence}, which pertains to the sensing modality and the representation of the signals. The crucial observation is that one can design efficient sensing or sampling protocols that capture the useful information content embedded in a sparse signal and condense it into a small amount of data. These protocols are nonadaptive and simply require correlating the signal with a small number of fixed waveforms that are incoherent with the sparsifying basis. CS is a very simple and efficient signal acquisition protocol which sample at a low rate and later uses computational power for reconstruction from what appears to be an incomplete set of measurements.

\subsection{The sensing problem}
We suppose that information about a signal $f(t)$ is obtained by linear functionals
recording the values
\begin{equation}\label{eq_1}
    y_k = \langle f, \phi_k \rangle
\end{equation}
with $k=1,\ldots,m$. That is, we simply correlate the object we wish to acquire with the waveforms $\phi_k(t)$. Although one could develop a CS theory of continuous time/space signals, we restrict our attention to discrete signals ${\mathbf f} \in  {\mathbf R}^n$ and a sensing matrix ${\mathbf \Phi} = [\phi_1,\phi_2,\ldots,\phi_n] \in  {\mathbf R}^{n \times n}$. We are then interested in undersampled situations in which the number $m$ of available measurements is much smaller than the dimension $n$ of the signal ${\mathbf f}$. Letting ${\mathbf \Phi}_s $ denote the $m \times n$ sensing matrix with the vectors ${\mathbf \phi}_1^* , \ldots , {\mathbf \phi}_m^*$ as rows ($a^*$ is the complex transpose of $a$), the process of recovering ${\mathbf f} \in {\mathbf R} ^n$ from
\begin{equation}\label{eq_1a}
{\mathbf y} = {\mathbf \Phi_s} {\mathbf f} \in {\mathbf R}^m
\end{equation}
is ill-posed in general when $m < n$, since there are infinitely many candidate signals $\widetilde{{\mathbf f}}$ for which ${\mathbf \Phi}_s \widetilde{{\mathbf f}} = {\mathbf y}$. But one could imagine a way out by relying on realistic models of objects ${\mathbf f}$ which naturally exist.

\subsection{Sparse representation}
Many natural signals have concise representations when expressed in a convenient basis. Mathematically speaking, we have a vector ${\mathbf f} \in {\mathbf R}^n$ which we expand in an orthonormal basis ${\mathbf \Psi} = [{\mathbf \psi}_1 , \ldots , {\mathbf \psi}_n]$ (compressed basis) as follows:
\begin{equation}\label{eq_2}
    {\mathbf f} = \sum_{i=1}^{n} x_i {\mathbf \psi}_i = {\mathbf \Psi} {\mathbf x}
\end{equation}
where ${\mathbf x} = [x_1,\ldots,x_n]^T$ is the representation of ${\mathbf f}$ respect to the basis  ${\mathbf \Psi}$. If most of the components of $ {\mathbf x}$ are zero, then ${\mathbf x}$ is referred to as a sparse representation of ${\mathbf f}$, and ${\mathbf \Psi}$ is a sparsifying basis. It is clear that from a signal with a sparse expansion one can discard the small coefficients without much perceptual loss. Now we consider the pair $({\mathbf \Phi}, {\mathbf \Psi})$ of orthobases of ${\mathbf R}^n$. The first basis ${\mathbf \Phi}$ is used for sensing the object ${\mathbf f}$ as in the equations \ref{eq_1} and \ref{eq_1a}, and the second is used to represent ${\mathbf f}$. The coherence $\mu({\mathbf \Phi}, {\mathbf \Psi})$ measures  the largest correlation between any two elements of ${\mathbf \Phi}$ and ${\mathbf \Psi}$. CS is mainly concerned with low coherence pairs. In fact, it is demonstrated that selecting $m$ measurements in the ${\mathbf \Phi}$ domain uniformly at random, the smaller the coherence the fewer $m$ samples are needed and one suffers no information loss by measuring just about any set of $m$ coefficients \cite{Candes,Donoho}. Since, in our case,  ${\mathbf \Phi}$ is the identity matrix and  ${\mathbf \Psi}$ is the Discrete Cosine Transform (DCT) basis, than a maximal incoherence is obtained. Moreover, the $m$ rows of the ${\mathbf \Phi_s}$ matrix are randomly selected in the ${\mathbf \Phi}$ domain. 

\subsection{Undersamplig and sparse signal recovery}
In a general signal processing problem we would like to measure all the $n$ coefficients of ${\mathbf f}$, but we only get to observe a subset of these and collect the data as in equation \ref{eq_1a}.

\noindent With this information, the signal is recovered by setting an $L_1$-norm constrained minimization problem; the proposed reconstruction ${\mathbf f}^*$ is given by
${\mathbf f}^* = {\mathbf \Psi} {\mathbf x}^*$, where ${\mathbf x}^*$ is the solution to the convex optimization program $(\|{\mathbf x}\|_{L_1} = \sum_i |x_i|)$
\begin{equation}\label{eq_4}
\begin{array}{ccccc}
    \mbox{ min } & \| {\mathbf x} \|_{L_1} & \mbox{ subject to } & y_k = \langle \phi_k, \Psi {\mathbf x} \rangle  & \mbox{        }  \forall k \in M\\
    {\mathbf x} \in R^n & & & &  \\
\end{array}
\end{equation}
That is, among all objects ${\mathbf f} = {\mathbf \Psi} {\mathbf x}$ consistent with the data, we pick the one with minimal $L_1$-norm.

\noindent The keys to CS are sparsity and the $L_1$ norm. If the expansion of the original signal as linear combination of the selected basis functions has many zero coefficients, then it is often possible to reconstruct the signal exactly (see \cite{Candes,Donoho} for more details and proofs). In principle, computing this reconstruction should involve the $L_0$ norm of ${\mathbf x}$, i.e., the number of its non-zero components. This is a combinatorial problem whose computational complexity is NP-hard. Fortunately, in \cite{Donoho} and \cite{Candes} the authors have shown that $L_0$ can be replaced by $L_1$.


\section{Signal Reconstruction} \label{sec_3}
\begin{figure}[tp]
\centering
\includegraphics[width=5in,keepaspectratio]{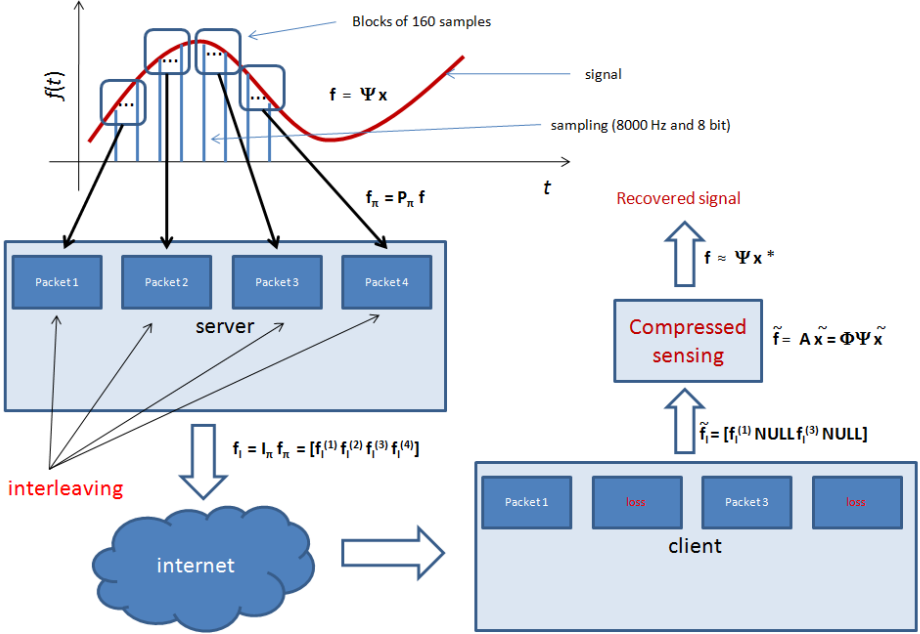}
\caption{Multimedia streaming process.}
\label{fig_3}
\end{figure}

To explain the proposed methodology we consider a multimedia streaming schema as shown in Figure \ref{fig_3}. In particular a client receives packets from a server. On the server a signal $f(t)$ is sampled by a PCM encoding technique (i.e., $64$ Kbit/s).  The server collect data each $20$ ms obtaining $4$ packets composed by $160$ bytes  (or $160$ samples). Before to apply the interleaving approach the data are randomly permuted to ensure a random distribution of the missing information.  In details, a raw signal can be regarded as a vector ${\mathbf f}$ that can be represented as a linear combination of certain basis functions as in equation \ref{eq_2}
\begin{equation}\label{eq_6}
     {\mathbf f} =  {\mathbf \Psi} {\mathbf x}
\end{equation}
The basis functions must be suited to a particular application (e.g. Wavelet, Gammatone, $\ldots$) and in our experiments, ${\mathbf \Psi}$ is the DCT. Notice that in order to use DCT as sparsifying basis, we have to rely on a moderately low number of samples ($640$ samples, $20$ ms). Next step of the process is the random permutation of the components of ${\mathbf f}$. If we suppose to have a random permutation matrix ${\mathbf P}_{\pi}$ then the resulting sequence is
\begin{equation}\label{eq_6a}
     {\mathbf f}_{\pi} =  {\mathbf P}_{\pi}{\mathbf f}.
\end{equation}
Applying the interleaving technique (permutation matrix ${\mathbf I}_{\pi}$), from the sequence ${\mathbf f}_{\pi}$ a new sequence of $4$ blocks ${\mathbf f}_{I}^{(i)}$, with $i=1,2,3,4$ is obtained
\begin{equation}\label{eq_6b}
     {\mathbf f}_{I} =  {\mathbf I}_{\pi}{\mathbf f}_{\pi} = [{\mathbf f}_{I}^{(1)}  \mbox{  } {\mathbf f}_{I}^{(2)} \mbox{  } {\mathbf f}_{I}^{(3)} \mbox{  } {\mathbf f}_{I}^{(4)}].
\end{equation}
Now we could consider that in a streaming communication process some packets may be lost (for example $2$ lost packets in Figure \ref{fig_3}). In this case the client receives only two packets
\begin{equation}\label{eq_6c}
     \widetilde{{\mathbf f}}_{I}  =  [{\mathbf f}_{I}^{(1)}  \mbox{  } Null  \mbox{  } {\mathbf f}_{I}^{(3)} \mbox{  } Null].
\end{equation}
The client applies the inverse of the interleaving process obtaining a subset of coefficients of ${\mathbf f}_{\pi}$
\begin{equation}\label{eq_6d}
     \widetilde{{\mathbf f}}_{\pi}  = {\mathbf I}_{\pi}^{T}\widetilde{{\mathbf f}} _{I}.
\end{equation}
Moreover, applying the inverse of the permutation process, the following subset of samples of ${\mathbf f}$ are obtained
\begin{equation}\label{eq_6e}
     \widetilde{{\mathbf f}}  = {\mathbf P}_{\pi}^{T}\widetilde{{\mathbf f}} _{\pi}.
\end{equation}
We could note that, in this way, the signal received by the client is a vector containing few random samples of ${\mathbf f}$  (not {\it Null} elements of $\widetilde{{\mathbf f}}$). Mathematically, we can consider a linear operator ${\mathbf \Phi_s} $ (as in equation \ref{eq_1a}) such that
\begin{equation}\label{eq_7}
     \widetilde{{\mathbf f}} =  {\mathbf \Phi_s} {\mathbf f}.
\end{equation}
In our case, ${\mathbf \Phi_s}$ is a random subset of the rows of the identity operator (i.e., the matrix ${\mathbf \Phi}^T$) with positions corresponding at the not {\it Null} elements of $\widetilde{{\mathbf f}}$.  To reconstruct the signal, the client recovers the sparse representation coefficients by solving the undetermined linear system 
\begin{equation}\label{eq_8}
     {\mathbf A}  {\mathbf x}=   \widetilde{{\mathbf f}}
\end{equation}
where ${\mathbf A} = {\mathbf \Phi}{\mathbf \Psi}$ is the compressive sensing matrix, i.e., computing the solution ${\mathbf x}^*$ to the convex optimization problem in equation \ref{eq_4}. Once we have the sparse representation of ${\mathbf x}$, we can recover the signal itself by computing
\begin{equation}\label{eq_9}
     {\mathbf f}  = {\mathbf \Psi} {\mathbf x}^*. 
\end{equation}

\section{Experimental Results} \label{sec_4}
\begin{figure}[tp]
\centering
\includegraphics[width=5in,keepaspectratio]{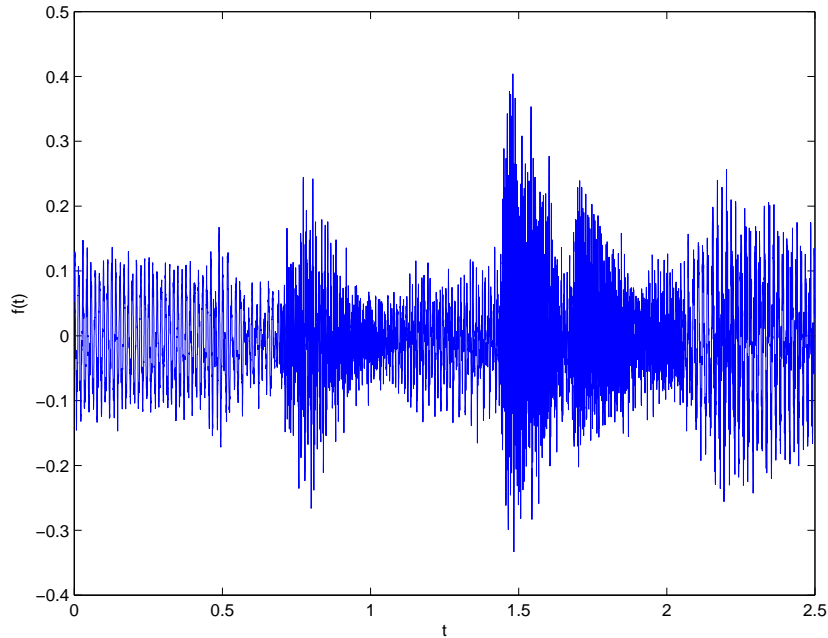}
\caption{Audio signal of a female speaker.}
\label{img_exp_1}
\end{figure}

\begin{figure}[tp]
\centering
\includegraphics[width=5in,keepaspectratio]{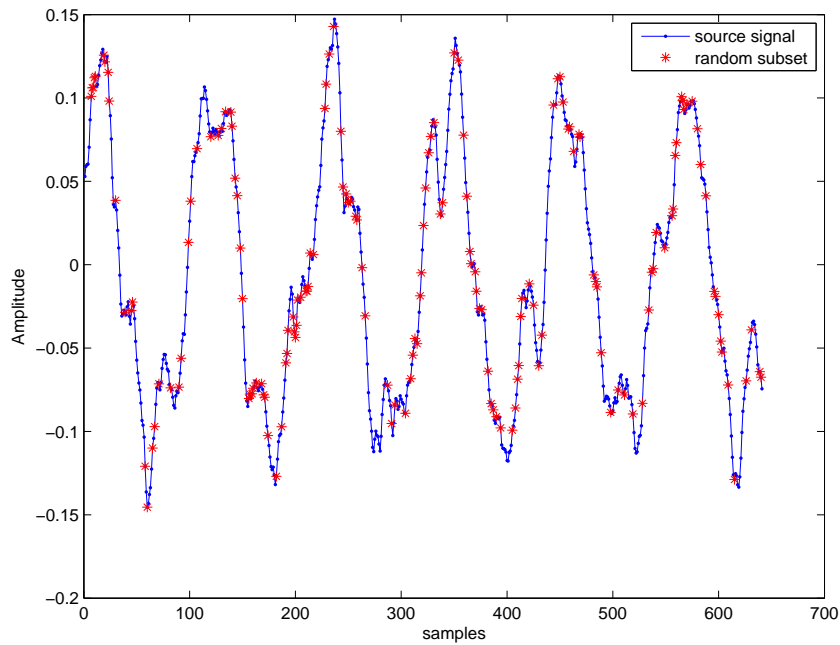}
\caption{Frame information after $3$ packets lost.}
\label{img_exp_2}
\end{figure}

\begin{figure}[tp]
\centering
\includegraphics[width=5in,keepaspectratio]{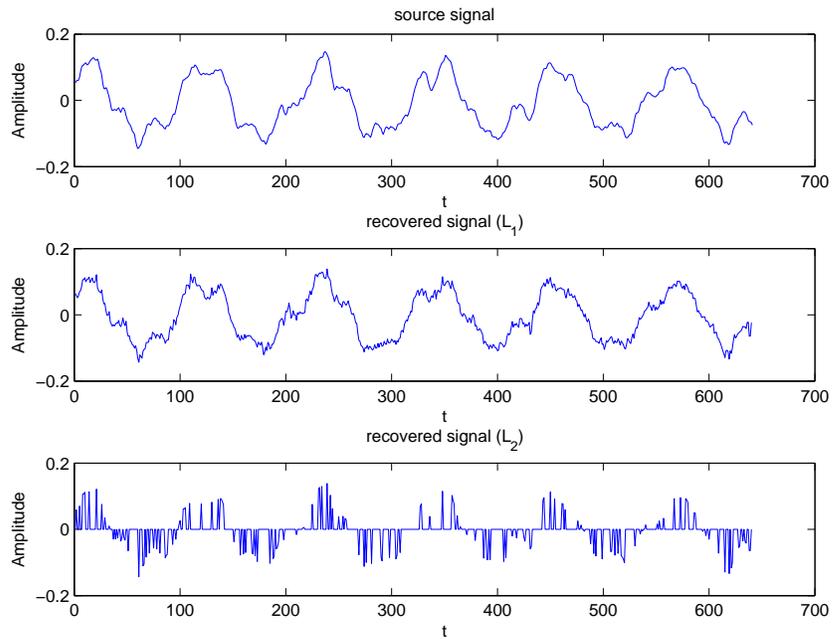}
\caption{Comparison between a frame of the source signal and those of the recovered signals by using $L_1$ and $L_2$ norms.}
\label{img_exp_3}
\end{figure}

\begin{figure}[tp]
\centering
\includegraphics[width=5in,keepaspectratio]{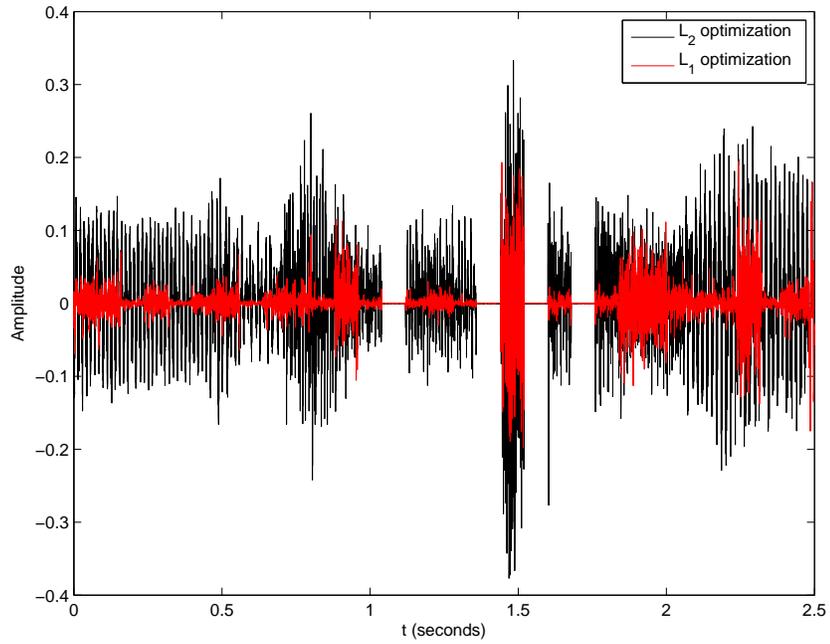}
\caption{Resisdua between the source signal and the recovered signals by using $L_1$ and $L_2$ norms.}
\label{img_exp_4}
\end{figure}

\begin{figure}[tp]
\centering
\includegraphics[width=5in,keepaspectratio]{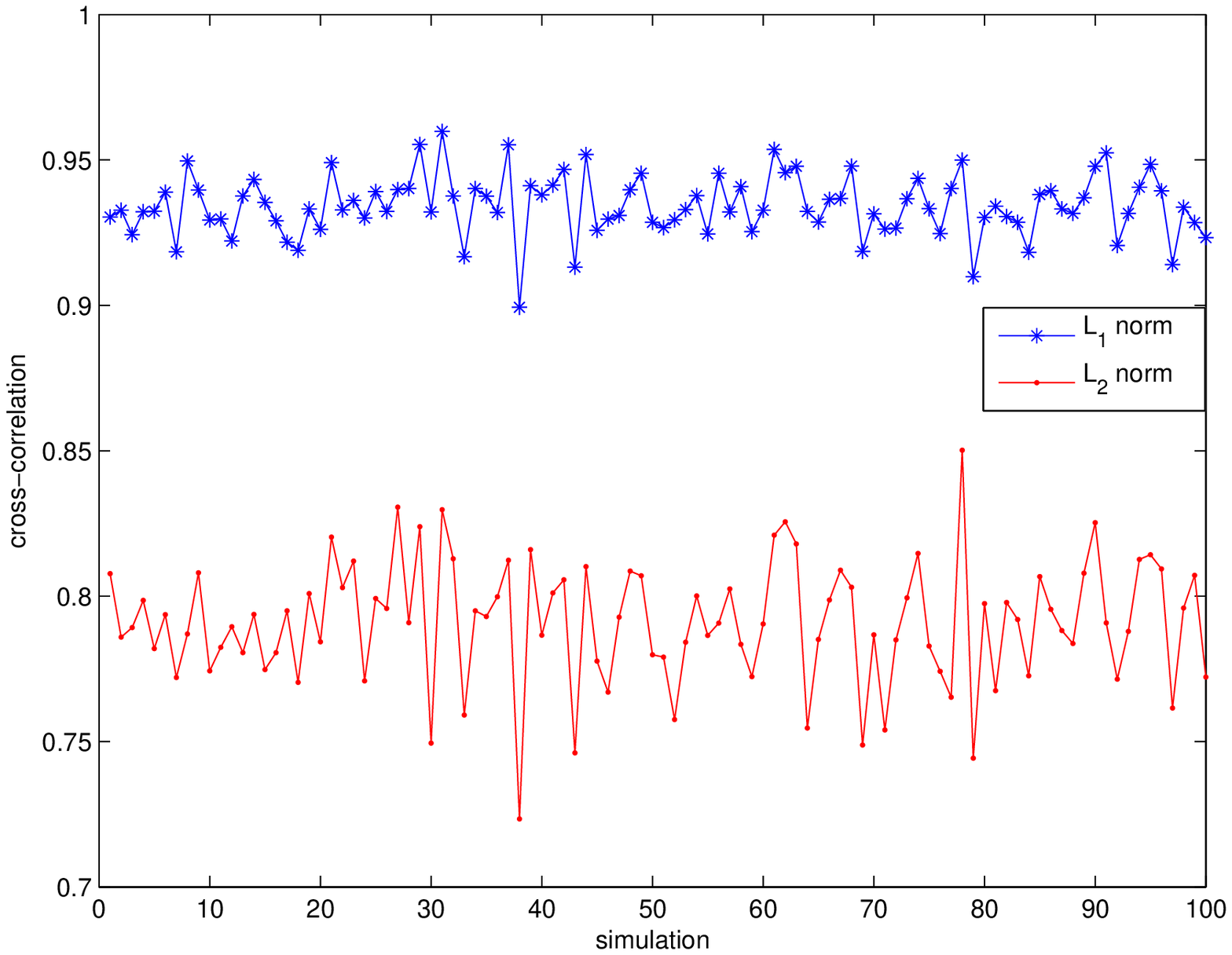}
\caption{Cross-correlation coefficients after $100$ simulations: audio female speaker. }
\label{img_exp_5}
\end{figure}

In this section we show some experimental results obtained applying the proposed methodology for reconstructing streaming audio signals. In the following we consider audio signals codified by a PCM encoding schema (sampling frequency of $8000$ Hz and $8$ bits of quantization). The results are presented comparing the source signals with those obtained from the optimization approaches by using both $L_1$ and $L_2$ norms, respectively. The first audio corresponds to a female voice that reads news in English. In Figure \ref{img_exp_1} we show a particular of this audio signal. In Figure \ref{img_exp_2} a frame of the signal ($\widetilde{{\mathbf f}}$ in equation \ref{eq_7}) after the interleaving phase and the loss of $3$ packets, is shown. In Figure \ref{img_exp_3} we compare a source frame of this audio signal with those recovered by using both $L_1$ and $L_2$ norms and when the loss of $3$ packets is considered. In this case the parameters $n$ and $m$ of the CS schema are $640$ and $160$, respectively. In Figure \ref{img_exp_4} the residua between the entire source signal and the recovered ones are visualized,  when the random loss of $0,1,2$ or $3$ packets is simulated. Finally, for this signal, we compare the correlation coefficients between the entire source signal and the recovered ones simulating the random loss of $0,1, 2$ or $3$ packets for each interleaving block.  The results are presented on $100$ simulations as presented in Figure \ref{img_exp_5}. 

\begin{figure}[tp]
\centering
\includegraphics[width=5in,keepaspectratio]{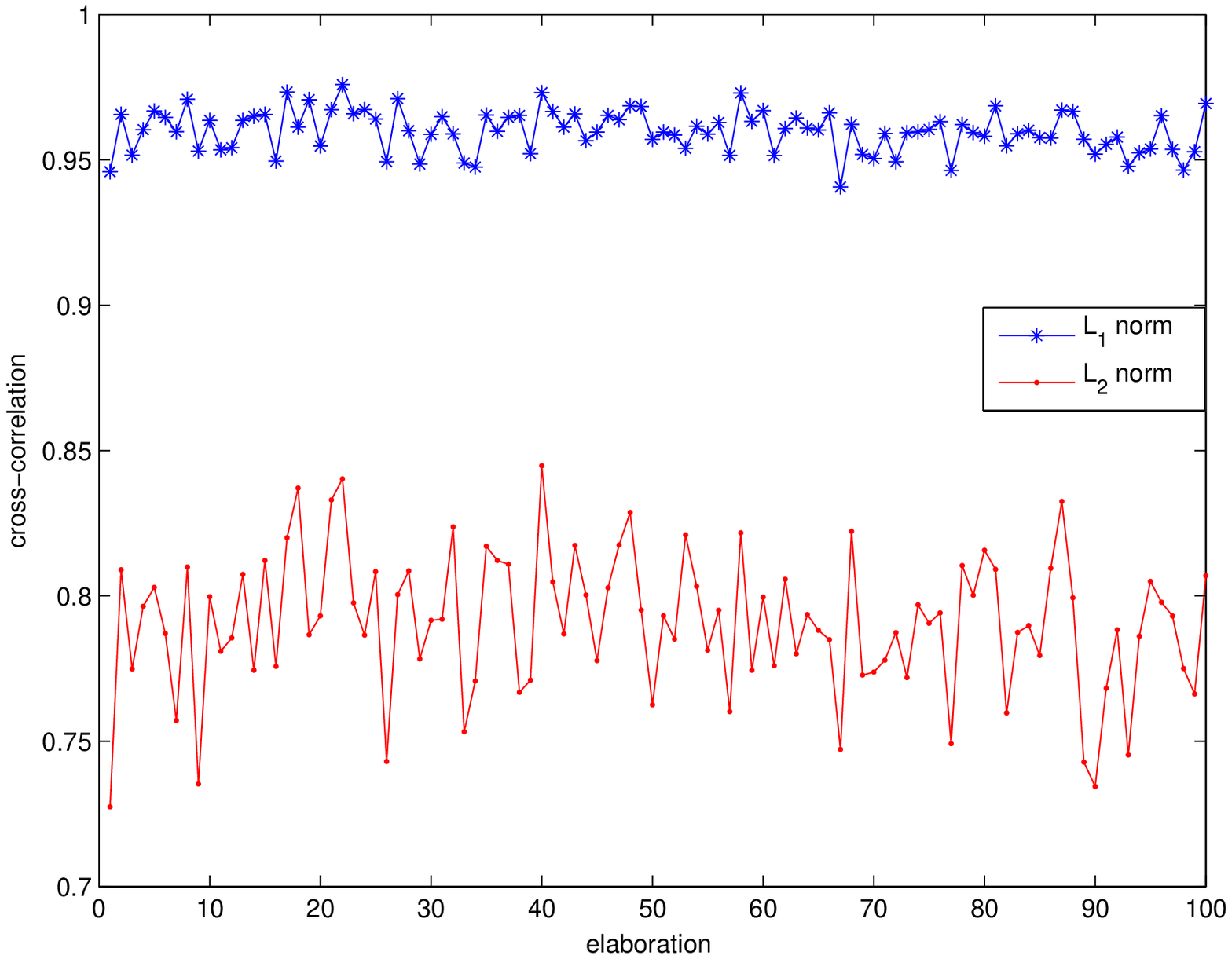}
\caption{Cross-correlation coefficients after $100$ simulations: audio male speaker.}
\label{img_exp_6}
\end{figure}

\noindent In the second experiment the audio corresponds to a male voice that reads news in English. As in the previous case, in Figure \ref{img_exp_6} the correlation coefficients obtained after $100$ simulations are shown. 

\begin{figure}[tp]
\centering
\includegraphics[width=5in,keepaspectratio]{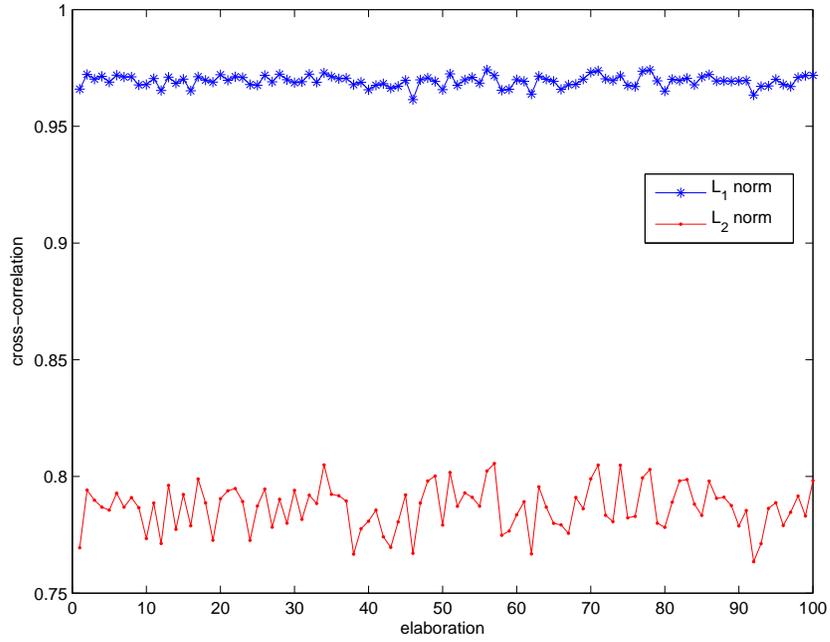}
\caption{Cross-correlation coefficients after $100$ simulations: audio song.}
\label{img_exp_7}
\end{figure}

\noindent In the last experiment we consider a Jazz audio song played by Chet Baker, titled ``Blue Room''. In Figure \ref{img_exp_7} the results of the correlation coefficients are presented. We can observe that in all the cases the $L_1$ norm permits to obtain the best results.

\section{Conclusions} \label{sec_5}
In this paper a new schema for data loss recovery, based on a Compressive Sensing technique, in multimedia streaming has been introduced. The audio streaming data are encapsuled in different packets by using an interleaving technique. Information contained in the loss packets is recovered by using a Compressive Sampling technique based on a $L_1$ norm. The experimental results highlighted that in the optimization schema $L_1$ norm perform better than $L_2$ norm. In the next future the authors will focus on the use of different optimization approaches and realization of the proposed schema for real applications (e.g. voice over ip, conferences, wireless streaming communications, $\ldots$), also in the case of dedicated hardware.

\end{document}